# The Interacting Energy Bands of Magic Angle Twisted Bilayer Graphene Revealed by the Quantum Twisting Microscope

J. Xiao[1†], A. Inbar[1†], J. Birkbeck[1†], N. Gershon[1], Y. Zamir[1], T. Taniguchi[2], K. Watanabe[2], E. Berg[1] and S. Ilani[1*]

[1] *Department of Condensed Matter Physics, Weizmann Institute of Science, Rehovot 76100, Israel.*
[2] *National Institute for Materials Science, 1-1 Namiki, Tsukuba, 305-0044 Japan.*
[†] These authors contributed equally to the work.
[*] Correspondence to: shahal.ilani@weizmann.ac.il

Electron interactions in quantum materials fundamentally shape their energy bands and, with them, the material's most intriguing quantum phases. Magic angle twisted bilayer graphene (MATBG) has emerged as a model system, where flat bands give rise to a variety of such phases, yet the precise nature of these bands has remained elusive due to the lack of high-resolution momentum-space probes. Here, we use the quantum twisting microscope (QTM) to directly image the interacting energy bands of MATBG with unprecedented momentum and energy resolution. Away from the magic angle, the observed bands closely follow the single-particle theory. At the magic angle, however, we observe bands that are completely transformed by interactions, exhibiting light and heavy electronic character at different parts of momentum space. Upon doping, the interplay between these light and heavy components gives rise to a variety of striking phenomena, including interaction-induced bandwidth renormalization, Mott-like cascades of the heavy particles, and Dirac revivals of the light particles. We also uncover a persistent low-energy excitation tied to the heavy sector, suggesting a new unaccounted degree of freedom. These results resolve the long-standing puzzle in MATBG – the dual nature of its electrons – by showing that it originates from electrons at different momenta within the same topological heavy fermion-like flat bands. More broadly, our results establish the QTM as a powerful tool for high-resolution spectroscopic studies of quantum materials previously inaccessible to conventional techniques.

Since the discovery of MATBG[1–3] a central challenge has been to determine whether its quantum states are best understood through a local framework or a topological one. On one hand, its correlated insulating states[2,4–8] observed at integer moiré fillings resemble Mott insulators, which are naturally described by localized electrons. On the other hand, its competing Chern insulating states[9–18] can only be explained through topological arguments. Perhaps the most puzzling observation, however, is that electrons in MATBG appear to exhibit contradictory behaviors even within the same electronic state: Compressibility[19] and spectroscopic[20] measurements have revealed cascades of Dirac-like compressible states, characteristic of itinerant electrons. Yet, entropy measurements[21,22] have uncovered a giant magnetic entropy that could only be explained if nearly all electrons behave as local moments. This striking duality[21]—simultaneous signatures of both localized and extended behavior—remains one of the most fundamental unresolved puzzles in the study of MATBG.

A range of theoretical approaches, including Hartree[23–25], Hartree-Fock[19,26–29], Quantum Monte Carlo[30,31], exact diagonalization[32,33], density matrix renormalization group (DMRG)[34–36] and dynamical mean-field theory (DMFT)[37,38], showed that interactions can profoundly affect the flat bands predicted by the Bistritzer-MacDonald (BM) model[1]. A key insight, solidified with the development of the Topological Heavy Fermion[39–41] and Mott Semimetal[42] frameworks, is that the topological nature of MATBG's flat bands may give rise to its electronic duality: while the electrons largely behave as heavy, localized "$f$-electrons", topology mandates the existence of momentum space regions where they act instead as light, delocalized "$c$-electrons". These two distinct electronic characters are thus predicted to coexist within the same flat bands, occupying different regions of momentum space. Although nano-ARPES experiments have provided rough evidence for the presence of flat bands[43–47], the absence of a technique capable of resolving their detailed momentum-space structure has, so far, left these fundamental puzzles unanswered.

In this work we present high-resolution momentum-space images of the interacting energy bands of MATBG, obtained using quantum twisting microscopy at $T = 4K$ [48]. Away from the magic angle, the observed bands closely follow the BM prediction, displaying Dirac points at the mini Brillouin zone (BZ) corners with renormalized Fermi velocity. At the magic angle, however, the bands are dramatically reshaped by interactions: across most of momentum space, the bands become extremely flat and largely gapped, hallmark of heavy *f*-electrons behavior. In contrast, near the Γ point of the mini-BZ, the bands are dispersive and gapless, consistent with the presence of light *c*-electrons. The evolution of the band structure with carrier density reveals Mott-like cascades of the heavy electrons and Dirac-like compressibility dominated by the light electrons. We directly observe how the reshuffling of charge between these extended and localized states underlies the previously reported Dirac revivals[19,20]. While resolving the long-standing dual-nature puzzle, our measurement also uncovers a new ~15meV excitation, not captured by current models, hinting at an important, yet unaccounted-for, degree of freedom in MATBG.

**Experimental Setup**

Our experiment probes momentum-resolved tunneling across a two-dimensional junction formed between monolayer graphene (MLG) on the QTM tip (purple, Fig. 1a) and twisted bilayer graphene (TBG) on a flat substrate (red/blue, Fig. 1a). The TBG device includes a metallic back gate and is capped with bilayer $WSe_2$ tunneling barrier (green, Fig. 1a). The experiment has two relevant angles: $\theta_{TBG}$, the TBG's twist angle, and $\theta_{QTM}$, the angle between the MLG on the tip and top layer of the TBG. The QTM enables continuous tuning of $\theta_{QTM}$ during the experiment with millidegree precision. While $\theta_{TBG}$ is fixed at any location in the sample, it naturally varies from one location to another due to inherent twist-angle disorder[49], allowing us to explore a range of $\theta_{TBG}$ values by performing experiments at different spatial locations.

In momentum space (Fig. 1b), the TBG's energy bands reside within a mini-BZ (black hexagon) whose corners, $K_T$ and $K_B$ (red/blue), correspond to the K points of its top and bottom graphene layers. By rotating the QTM, we rotate the Dirac point of the MLG probe (purple circle) along an arc in momentum space (dashed purple line) that passes through the $K_T$ and $K_B$ points of the TBG, and closely approaches the two Γ points in adjacent Brillouin zones, enabling us to map the TBG energy bands along the $\Gamma - K_T - M - K_B - \Gamma$ trajectory ("QTM trajectory").

Fig. 1c presents the canonical single-particle BM energy bands of TBG along the QTM trajectory for a twist angle slightly larger than the magic angle ($\theta_{TBG} = 1.2°$). Theory predicts that the two low-energy flat bands will exhibit symmetry-protected Dirac points at $K_T$ and $K_B$ and reach the maximal width at the Γ points, where energy gaps separate them from the higher-energy remote bands. The bands are colored by the weight in momentum space projected on the top TBG layer, to which the electrons tunnel.

**The energy bands at $\theta_{TBG} = 1.2°$**

We begin with QTM measurements of TBG with $\theta_{TBG} = 1.2°$. Fig. 1d shows the tunneling $dI/dV$ measured as a function of $\theta_{QTM}$ and bias voltage, $V_b$, at moiré filling $\nu \approx 0$. Recalling[48,50] that $\theta_{QTM}$ and $V_b$ translate to momentum and energy via $k = \theta_{QTM} K_D$ ($K_D$ is the Dirac momentum) and $E = eV_b$, this measurement directly maps the TBG's energy bands along the QTM trajectory (the corresponding k-points are indicated on the top axis). The positions of the peaks in $dI/dV$ along the $V_b$ axis reveal the energy of the electronic bands, while the peaks' amplitude provide insight into the wavefunctions. To better trace the energies of the flat bands, for each $\theta_{QTM}$ we normalize the valence flat band peak to unity and plot the resulting $dI/dV_{norm}$ in Fig. 1e. The very high resolution in the measurement ($\delta k \sim 0.1° \cdot K_D$ and $\delta E \leq 10 meV$, Methods M3) allows us to visualize these bands with unprecedented details. Specifically, we observe a Dirac dispersion with a crossing at $K_T$ but with a Fermi velocity that is seven times smaller than that in monolayer graphene. At Γ, the flat bands attain their maximum width of 70meV, separated by gaps of 50meV and 40meV from the remote conduction and valence bands, respectively. Moreover, we can also identify van Hove singularities in the remote bands at energies of E=174meV and -171meV.

Fig. 1f presents a theoretical calculation of the momentum-resolved $dI/dV$, using the non-interacting BM bands at $\theta_{TBG} = 1.2°$, considering the tunneling matrix elements as well as the electrostatics of the MLG-TBG junction and the Dirac dispersion of the MLG probe (Methods M5). Visibly, for this larger-than-magic-angle TBG, the single-particle model captures well both the band energies and the momentum-dependent tunneling amplitudes.

**The energy bands at the Magic Angle**

We now turn to the central results of our paper – the energy bands of magic angle TBG. The measurement is performed in a region of the sample where room-temperature conducting AFM measurements observed a regular moiré pattern with a periodicity of approximately 13nm (Methods M6) corresponding to $\theta_{TBG} = 1.1°$. Fig. 2a displays $dI/dV$ measured with the QTM as a function of $\theta_{QTM}$ and $V_b$, at a carrier density close to charge neutrality ($\nu = -0.2$). Fig. 2b presents the normalized $dI/dV$ ($dI/dV_{norm}$), following the normalization described earlier.

Remarkably, the bands observed at $\theta_{TBG} = 1.1°$ are dramatically different from those at $\theta_{TBG} = 1.2°$. They do not show a dispersing Dirac point at $K_T$ but instead feature two extremely flat bands, separated by a large energy gap of approximately 35meV. Curiously, the gap appears at almost all momenta along the measured trajectory, except near the Γ point, where the bands are gapless. At this specific filling, the conduction flat band is significantly wider in energy than the valence band, signaling that it has a shorter electronic lifetime. Tracing the center energy of the bands vs. momentum (dots, Fig. 2c) shows that the bands are flat to within ~4meV. Compared to the bands predicted by the non-interacting BM-model for $\theta_{TBG} = 1.1°$ (gray lines), we see a drastic difference, demonstrating the central role of interactions in determining their shape.

In addition to revealing band energies, the spectral function also carries crucial amplitude information. Notably, we observe a gradual reduction in the $dI/dV$ amplitude of the flat bands along the momentum trajectory from $K_T$ to $K_B$ (Fig. 2a). To quantify the total spectral weight of a specific band – independent of its lifetime – we define the peak intensity, $I_{peak} = \int dI/dV \cdot dV_b$, obtained by integrating the area under its corresponding $dI/dV$ peak. Plotting $I_{peak}$ vs. $\theta_{QTM}$ for the valence flat band (Fig. 2d, dots), reveals a sharp increase near the left Γ point, followed by a gradual decline toward the $K_T$ point, and a steeper drop to near zero at $K_B$. Beyond $K_B$, the intensity remains negligible, falling below the measurement noise floor.

The magnitude of $dI/dV$ encodes tunnelling matrix element information, offering additional insights into the character of individual wavefunctions. Specifically, $dI/dV$ is proportional to $A(k,\omega) \cdot P_t(k) \cdot P_s(k)$, where $A(k,\omega)$ is the spectral function, $P_t(k)$ is the projection operator onto the top TBG layer, and $P_s(k) = \langle \sigma_x + I \rangle_k$ is the sublattice projection operator (Methods M5). At the magic angle, the BM model predicts that almost all wavefunctions in the mini-BZ have equal population on the top and bottom layers, $P_t(k) \sim 0.5$ (Methods M5). However, the sublattice projection should exhibit strong momentum dependence, which is highly sensitive to the ratio $w_0/w_1$, which reflects the relative size of AA and AB stacking regions. Comparing our measurements with theoretical predictions[51] for various $w_0/w_1$ ratios (colored traces), we find excellent agreement for $w_0/w_1 = 0.6$, while other values show significant discrepancies. This value represents a notable deviation from the accepted value of 0.8 and could have significant implications on the physics predicted in MATBG[36,52].

**Evolution of the Interacting Bands with Filling**

Fig. 3 presents the energy bands at all integer fillings from $\nu = -4$ to $\nu = 4$ as well as at $\nu = 4.6$. At most integer fillings, two sharp and extremely flat bands are observed. At charge neutrality ($\nu = 0$), these bands are symmetrically positioned about the Fermi energy, $E_F$ ($V_b = 0mV$). Upon electron doping, the conduction flat band becomes pinned at $E_F$ from above, while the valance flat band lies $\Delta E \approx 15 meV$ below $E_F$. The opposite occurs for hole doping, where the valence flat band becomes pinned to $E_F$ from below and the conduction flat band shifts to $\Delta E \approx 15 meV$ above it. By $\nu = 4.6$, $E_F$ has moved well into the remote bands.

While the bands shift in a largely rigid manner – maintaining their flatness across most of momentum space – the behavior near the Γ point is distinct. At $\nu = 0$ the bands at Γ are gapless and symmetrically positioned between the flat bands. However, with electron (or hole) doping, the bands undergo a pronounced stretching around Γ, with the Γ point gradually shifting to negative (or positive) energies. This shift reaches $W \sim 42 meV$ for $\nu = 4$ and $W \sim 54 meV$ for $\nu = -4$, leading to a highly renormalized band width upon doping.

The distinctive filling-dependence observed near the Γ point is closely linked to the predicted topological nature of the flat bands, which gives rise to fundamentally different wavefunctions near this momentum. Across most of the mini-BZ the flat bands' wavefunctions are expected to be highly localized at the AA moiré sites. In contrast, near the Γ point, the wavefunctions are predicted to become extended, with little to no charge density at the AA sites. In recent topological heavy fermion representations of these bands[39–41,53,54], the localized states correspond to "*f*-electrons", while the extended states near Γ correspond to itinerant "*c*-electrons". At charge neutrality, the electrostatic (Hartree) potential within a moiré unit cell is nearly uniform. However, as filling increases, electrons preferentially occupy the localized *f*-electron states, generating a repulsive Hartree potential that peaks at the AA sites. This repulsion affects the localized *f*-electrons more strongly than the itinerant *c*-electrons, introducing a Hartree-driven stretching of the flat bands, as seen in our experiment.

Further insight into the evolution of the electronic structure emerges when we examine partial moiré fillings. Fig. 3b, plots this evolution from $\nu = -0.9$ to $\nu = 0.9$, with dashed lines following the progression of key spectral features. The dashed red line tracks a flat band that appears initially at $\nu = -0.9$ at high energy as a diffuse "plume" — reflecting an incoherent band with a short lifetime. As filling increases, this band gradually moves toward $E_F$, increasing in spectral weight as it approaches. The dashed blue line follows a band with the opposite trend: it begins as a band with strong spectral weight just below $E_F$ and progressively shifts to large negative energies while becoming increasingly smeared. Notably, in both cases, just before the sharp band vanishes while crossing $E_F$ at $\nu \approx +1$ or $\nu \approx -1$, a new diffuse band appears at high energy. This behavior is most clearly seen in Fig. 3c, which presents the energy spectrum, momentum-averaged over the flat part of the bands (excluding the region near Γ, Methods M7), along with its decomposition into Gaussian contributions.

**Filling-Dependent Spectroscopy at Particular Momenta**

Recent topological heavy-fermions[39–41] and Mott-Semimetal[42] theories predict that the dual nature electrons in MATBG arises from the different character of the electronic states at different momenta within the flat bands. To investigate this, we leverage a key capability of the QTM - performing spectroscopy as a function of filling at a particular momentum.

Figure 4a presents $dI/dV$ measured as a function of $V_b$ and $\nu$ at momentum $K_T$. At charge neutrality, a clear energy gap is observed at this momentum. The evolution of the energy bands with filling – previously shown in Fig. 3b – appears here in the familiar form of electronic cascades[19,20]: near each integer $\nu$, an incoherent, high-energy band emerges, which gradually shifts with filling to lower energies, becoming stronger and sharper the closer it approaches $E_F$. This band crosses $E_F$ over a finite range of $\nu$, and once fully occupied, a new incoherent high-energy band emerges, while the previous one persists below $E_F$. This cascading behavior recurs at all integer fillings and is more pronounced on the electron-doped side. Identical filling-dependent spectral evolution is observed at all other momenta along the flat part of the bands (Methods M7). This explains why STM, which averages over all momenta, captures a similar pattern.

A markedly different behaviour is seen at the Γ point (Fig. 4b). Here, there is no apparent gap at neutrality, nor do we see any sign of cascading bands crossing through $E_F$ with filling. Instead, the $dI/dV$ displays a peak at $E_F$ at $\nu = 0$, which gradually shifts to negative energies with filling. The shift is not strictly monotonic but exhibits wiggles that align with integer fillings, more prominently seen on the electron-doped side. Interestingly, when we overlay the chemical potential, $\mu(n)$, obtained from the compressibility measurements of Zondiner *et al.* (black line)[19], we find that it closely follows this peak, capturing even its wiggles. Since that measurement was performed in a sample with a slightly larger twist angle ($\theta_{TBG} = 1.13°$) we do not expect perfect correspondence.

Zooming in further (Fig. 4c), we uncover a striking correlation between the behavior at the $K_T$ point, the Γ point, and the inverse compressibility, $d\mu/dn(n)$, from Zondiner *et al.*[19] (Fig. 4c top, center and bottom). From $\nu = 0$ to $\nu \approx 0.6$ the peaks at both Γ and $K_T$ shift monotonically to lower energies with increasing filling. At $\nu \approx 0.6$, as the flat band at $K_T$ reaches $E_F$, the peak at Γ reverses direction and begins to climb up in energy. Notably, this turning point coincides with a sign change in the total inverse compressibility, which becomes negative. This behavior continues until $\nu \approx 1$, where the flat band at $K_T$ becomes fully occupied, the inverse compressibility sharply rises, and the peak at Γ reverses direction again.

These observations can be naturally understood using the topological heavy fermion decomposition of the flat bands into $c$ and $f$ electrons, illustrated schematically in Fig. 4d. Within this basis, the $c$-electrons occupy doubly degenerate Dirac bands at the Γ point (cyan), while the $f$-electrons form flat bands (red). For simplicity, we neglect the $c$-$f$ hybridization. Within this picture, the measurement at Γ is tracking the filling-dependence of the $c$-electrons Dirac point, while that at $K_T$ follows the filling-dependence of the flat $f$-electron bands. At $\nu = 0$, the $f$-bands lie away from $E_F$ and the $c$-electrons Dirac point sits at $E_F$. Upon doping, carriers initially populate the $c$-bands, shifting their Dirac point downward in energy. Around a filling of $\nu \approx 0.6$ the $f$-bands reach $E_F$ and begin to populate. At this stage, carriers start transferring from the $c$-bands to the $f$-bands, depopulating the former and pushing their Dirac point back toward $E_F$. This is the underlying mechanism behind the "Dirac revivals"[19], only that it is not between flavors as originally postulated[19], but between $c$- and $f$-electrons. Visibly, there is a clear competition between the Fock-driven revivals that depopulate the $c$-bands with filling (upward wiggles), and the Hartree terms that progressively populate these bands with filling (overall downward slope) explaining why the revivals do not reset the $c$-bands fully to their neutrality point.

While some key aspects of our experiments are captured by theoretical models of interacting electrons in MATBG[26,27,30,31,37,38,41,42], one persistent feature remains unexplained: the excitation appearing at $\Delta E \approx +15 meV$ for holes and $\Delta E \approx -15 meV$ for electrons (green line in Figs. 3a and 4a). Similar features were previously observed in non-momentum-resolving STM measurements[17,20], but our momentum-resolved experiments now reveal that this excitation exists only in the flat section of the bands (Fig. 4a) and not at the Γ point (Fig. 4b), suggesting its connection with the $f$-electrons. However, unlike the $f$-electron energy levels, which vary strongly with filling, this excitation energy is independent of filling. The persistence of this feature across all measurements that exhibit cascading behaviour may hint at the presence of an intrinsic excitation or collective mode, whose origin is still unknown but may play a crucial role in the correlated and superconducting behaviour of MATBG.

**Doping Dependence of the Relative Fraction of c and f Electrons**

Finally, we show that the Hartree potential leads to a doping-dependent evolution of the fraction of *c*- and *f*- electrons within the mini-BZ. This effect is revealed through the zero-bias $dI/dV$, which reflects the spectral weight at $E_F$. When plotted as a function of $\theta_{QTM}$ and $\nu$ (Fig. 5), this spectral weight maps out where electronic states reside in momentum space as a function of doping. In this plot, the *f*-electron flat bands manifest as horizontal features spanning a wide range in momentum space. Their spectral weight oscillates with filling with a periodicity $\Delta \nu \approx 1$, consistent with the cascading behaviour of the *f*-electrons. The spectral weight of these features decays near the $\Gamma$ point as it is transferred to the *c*-electrons. Visibly, these features get progressively shorter with filling (highlighted by the white dashed line), indicating that the k-space region occupied by *f*-electrons shrinks while that occupied by *c*-electrons expands (illustrated on the right). The radius of the *c*-electron region, $r_c$, normalized by the diameter of the mini-BZ, $K_D$, $s = r_c/K_D$, is an important small parameter in a recent analytical theory of these bands[42]. Our measurements show that this parameter depends on filling: it starts small near neutrality and steadily increases with doping, reaching $s \approx 0.45$ at $\nu = 4$. For $\nu > 4$, the *c*-electron region is overtaken by the minimum of the remote bands, which appears as an intense $dI/dV$ near $\Gamma$ (see arrow).

Our measurements resolve a long-standing puzzle in the physics of MATBG: the dual nature of its electrons. By directly visualizing the energy bands, we show that this duality arises from distinct characteristics of the flat bands' wavefunctions at different regions of momentum space. Near the $\Gamma$ point the bands are dispersive, consistent with *c*-electron behavior, while elsewhere they remain extremely flat, characteristic of *f*-electrons. We directly observe how this leads to stretching of the bands due to Hartree potential, as well as to electronic cascades and "Dirac revivals" due to the reshuffling of charge between these two degrees of freedom. Qualitatively, our findings provide direct evidence for the topological heavy fermion and Mott semimetal descriptions of this system. However, we also uncover a new low-energy excitation, not accounted by any existing models, which may play an important role in the exotic physics of MATBG.

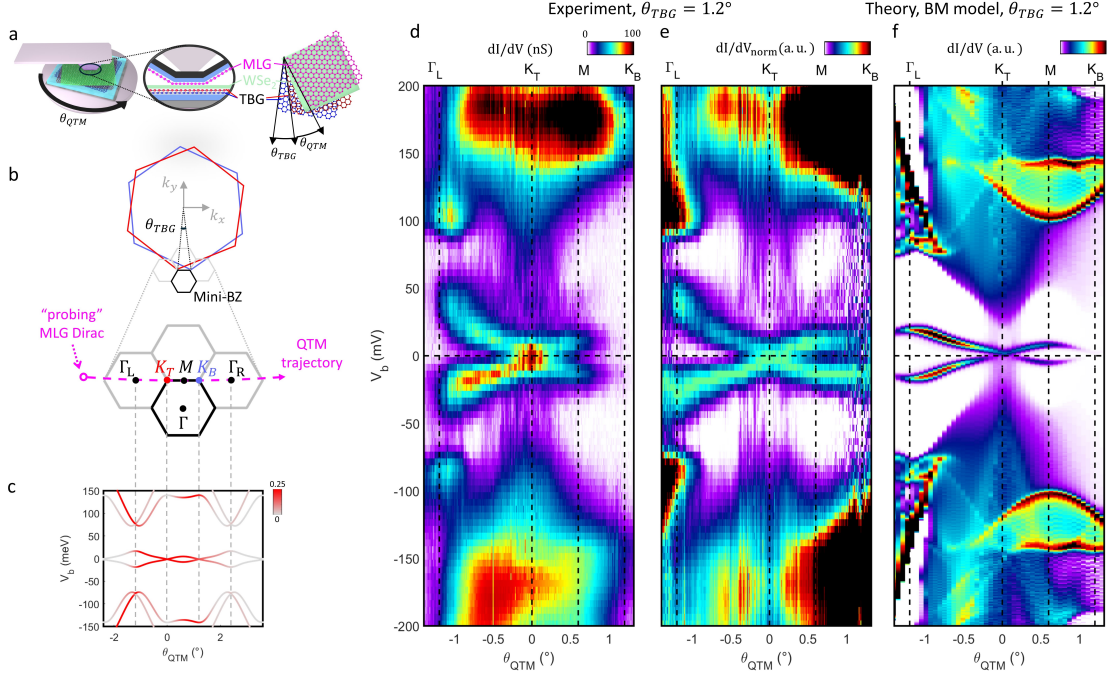

**Figure 1: Measurement setup and momentum resolved spectroscopy of twisted bilayer graphene (TBG) with $\theta_{TBG} = 1.2°$. a.** Schematic of the QTM measurement setup (left) and cross-sectional view (center), featuring a monolayer graphene (MLG, purple) on the QTM tip, separated by bilayer WSe$_2$ tunnel barrier (green) from a back-gated TBG (top and bottom layers in red and blue). We measure the differential tunneling conductance, $dI/dV$, between the MLG and TBG as a function of their relative bias voltage, $V_b$. The top view (right panel) defines the TBG twist angle, $\theta_{TBG}$, and the angle between MLG and top TBG layer, $\theta_{QTM}$. **b.** Momentum-space diagram showing the Brillouin zones (BZs) of the TBG's top (red) and bottom (blue) layers, the resulting mini-BZ (black), and neighboring mini-BZs (gray). Inset: When the tip is rotated with respect to the sample, the MLG's Fermi surface (purple circle) traces an arc in momentum space – the "QTM trajectory" (dashed purple) – intersecting the K points of the top and bottom TBG layers (K$_T$ and K$_B$) and closely approaching the Γ points in adjacent mini-BZs. **c.** Calculated single particle band structure using the non-interacting Bistritzer-Macdonald (BM) model[1] along the QTM trajectory for TBG with $\theta_{TBG} = 1.2°$. The flat bands feature Dirac points at K$_T$ and K$_B$ with renormalized Fermi velocity, while reaching their maximum bandwidth at the Γ points, where they are also minimally separated from the remote bands by energy gaps. The color scale indicates the wavefunction weight on the top TBG layer. **d.** Measured $dI/dV$ as a function of $\theta_{QTM}$ and $V_b$ for TBG with $\theta_{TBG} = 1.2°$ (see Methods M9). The top axis marks key momenta along the QTM trajectory: Γ, K$_T$, M and K$_B$. **e.** Same data as in panel d, but plotted as a normalized differential conductance, $dI/dV_{norm}$, where for each $\theta_{QTM}$ the peak intensity within the flat bands is normalized to unity. This allows us to follow the energy dispersion of the flat bands. **f.** Calculated momentum-resolved tunneling conductance using the non-interacting BM model[1] for TBG and incorporating the electrostatics of the MLG-TBG junction and the Dirac dispersion of the MLG probe (Methods M5). For this larger-than-magic-angle TBG, the single-particle model captures rather well the measured data.

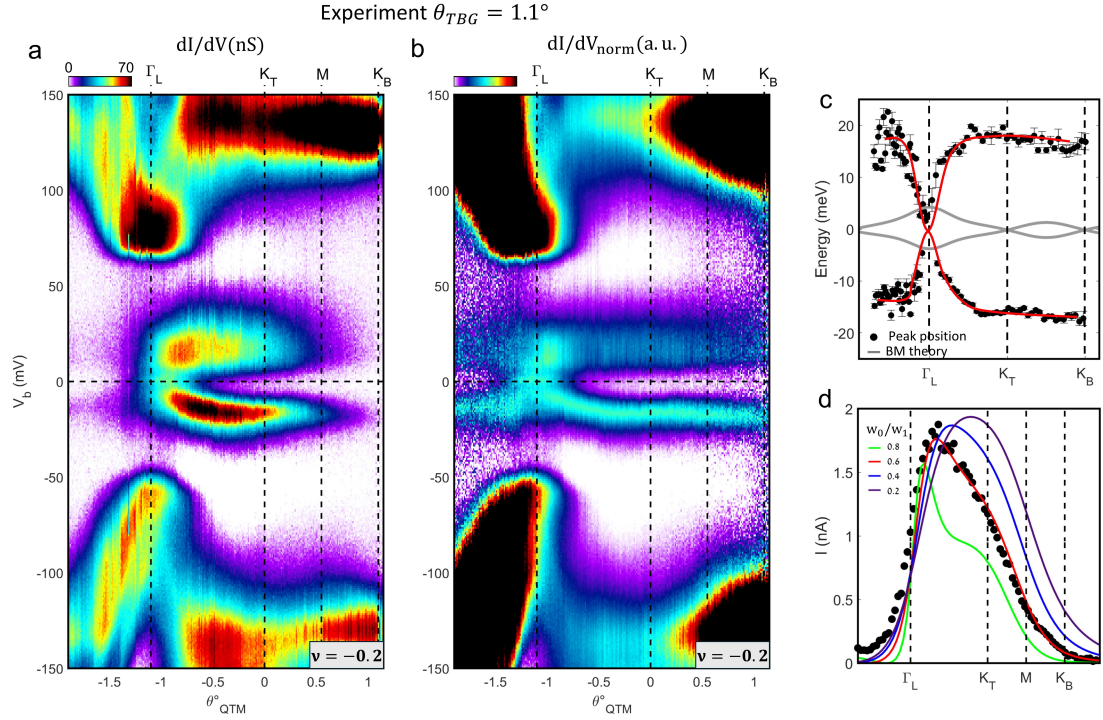

**Figure 2: Momentum resolved spectroscopy of Magic-Angle TBG. a**. Measured $dI/dV$ as a function of $\theta_{QTM}$ and $V_b$ at a filling factor of $\nu = -0.2$ for TBG with $\theta_{TBG} = 1.1°$. Key momenta ($\Gamma$, $K_T$, M and $K_B$) are marked on the top axis. **b.** Same data as in panel a, but for each $\theta_{QTM}$ we normalize the peak intensity of the valence flat band to unity, defining the quantity $dI/dV_{norm}$. **c.** Extracted flat band dispersion from panel a (dots), where we trace the center of the flat band peaks vs. $\theta_{QTM}$, with a guide to the eye (red). Gray lines plot the single particle BM bands for $\theta_{TBG} = 1.1°$. **d.** The intensity of the conduction flat band, given by integrating the area under its peak $I_{peak} = \int dI/dV \cdot dV_b$ in panel a, plotted as a function of $\theta_{QTM}$ (dots). We compare this to the theoretically predicted momentum-dependence of the intensity, given by $A(k,\omega) \cdot P_t(k) \cdot P_s(k)$, where $A(k,\omega)$ is the spectral function, $P_t(k)$ is the wavefunction projection on the top TBG layer and $P_s(k) = \langle \sigma_x + I \rangle_k$ is the sublattice projection. The calculation uses the wavefunctions of the BM model and the different lines correspond to a different $w_0/w_1$ ratios within this model (Methods M5). We observe an excellent fit for $w_0/w_1 = 0.6$.

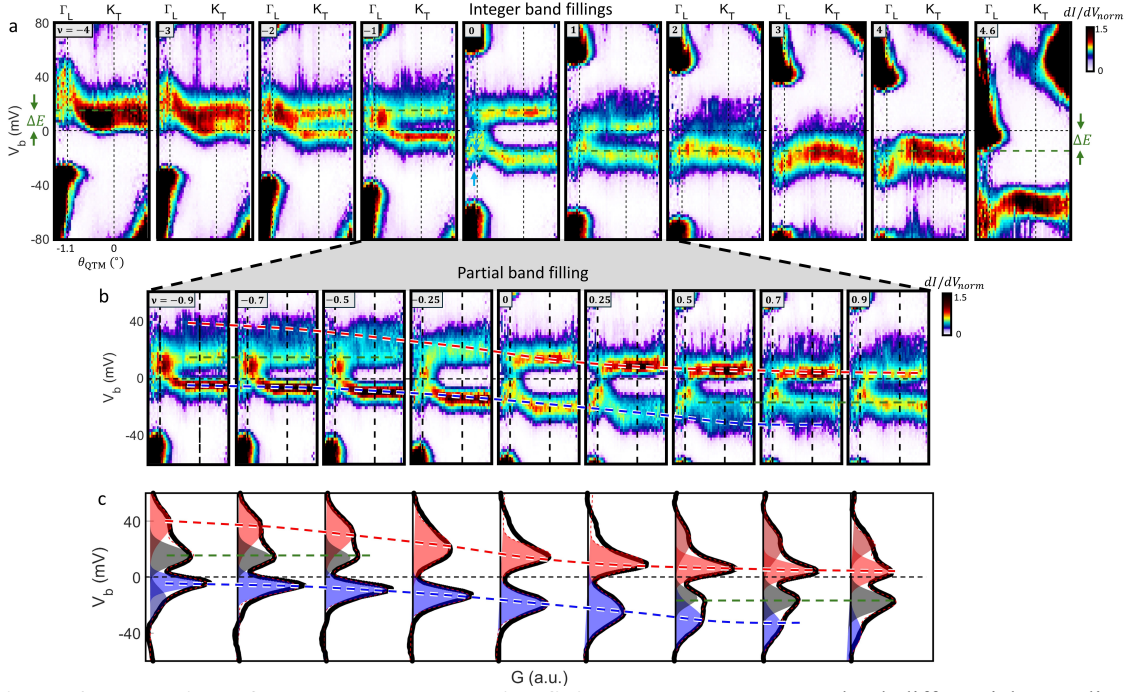

**Figure 3: Evolution of the energy bands with filling. a**. Momentum resolved differential tunneling, normalized as in Figs. 1e and 2b, $dI/dV_{norm}$, measured as a function of $\theta_{QTM}$ and $V_b$ for all integer fillings $\nu = -4$ to $\nu = 4$ and for $\nu = 4.6$ (the momenta $\Gamma$ and $K_T$ are indicated on the top x-axis). At all fillings the bands are flat throughout, except near the $\Gamma$ point. At neutrality, the two bands are symmetrically positioned around the Fermi level, $E_F$ ($V_b = 0$) and are separated by about 35mV, apart from at the $\Gamma$ point, where they touch. At integer electron doping, the conduction flat band is pinned at $E_F$ from above while the valance flat band lies $\Delta E \approx 15mV$ below $E_F$ (green dashed line). At integer hole doping the valence flat band becomes pinned to $E_F$ from below and the conduction flat band shifts to $\Delta E \approx 15meV$ above it (green dashed line). At $\nu = 4.6$, the flat bands are completely occupied and $E_F$ lies within the remote bands near the $\Gamma$ point. **b**. Measured $dI/dV_{norm}$ vs. $\theta_{QTM}$ and $V_b$ at fractional fillings between $\nu = -0.9$ and $\nu = 0.9$ (the error bar on the filling factor in this measurement is $\delta\nu \sim 0.1$). The dashed red line marks the evolution of one band from a high-energy smeared "plum" at negative doping to a narrow energy level pinned above $E_F$ at positive doping. The dashed blue line marks the opposite evolution of another band. The green dashed line marks two additional energy bands that coexist with the former and remain at fixed energies of $V_b = \pm 15meV$. **c**. Line cuts of $dI/dV$ vs. $V_b$ (black line), at the same filling factor as panel **b**. The line cuts are averaged over the flat band region (excluding the region near $\Gamma$, Methods M7). The line cuts are decomposed into different Gaussian contributions, marked by peaks with different colors, showing a multi-peak structure with different widths and amplitudes. The guiding lines in panel **b** are overlaid on top with the corresponding color of the Gaussian peaks. The sum of decomposed Gaussian contributions is plotted as the red line in each panel, showing a good fit.

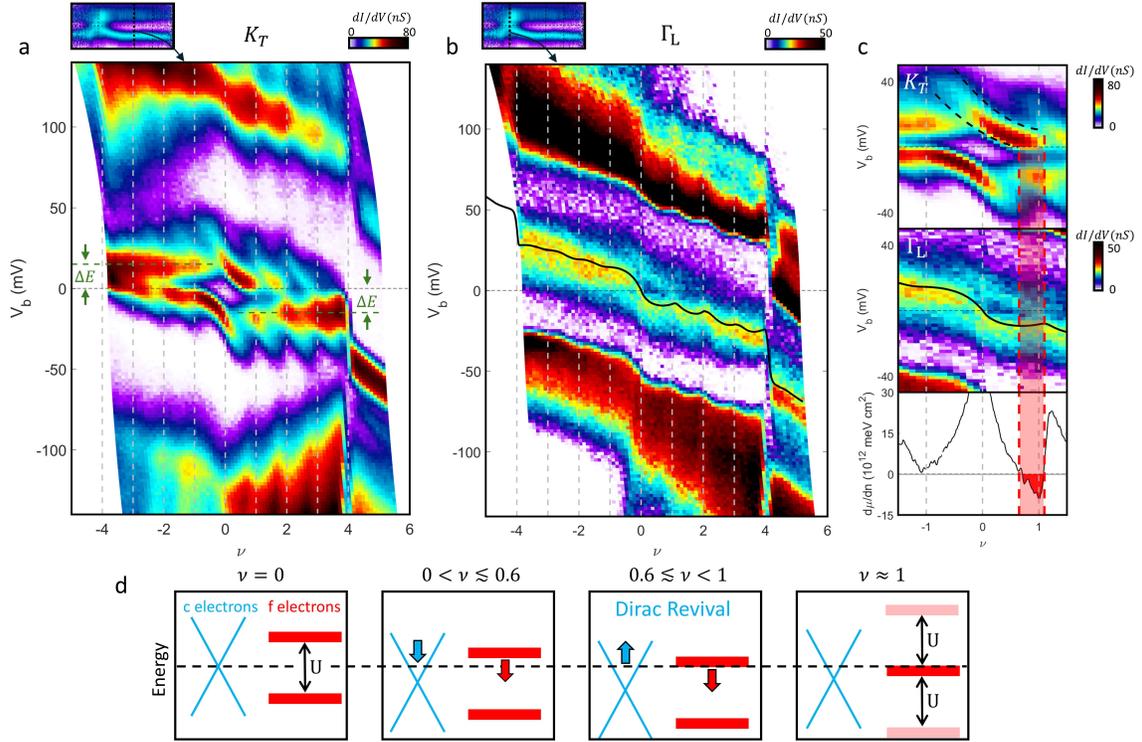

**Figure 4: Spectroscopy at fixed momentum points. a**. $dI/dV$ vs. $\nu$ and $V_b$ measured at momentum $K_T$ (top inset indicates this momentum on a zoom in of Fig. 2b). At this momentum, the data shows clear signatures of cascading behavior of the flat bands. Identical behavior is measured across all momenta along the flat bands (see extended Fig. 6). In addition to the cascades, fixed-energy excitations appear at $\Delta E \approx +15 meV$ for hole doping and $\Delta E \approx -15 meV$ for electron doping (green dashed lines). **b.** $dI/dV$ vs. $\nu$ and $V_b$ measured at the $\Gamma$ point (momentum location is indicated in the top inset). Overlaid is the chemical potential, $\mu(\nu)$ taken from the compressibility measurement in ref.[19] (black solid line), showing a surprising agreement between the filling dependence of the $dI/dV$ peak at the $\Gamma$ point and the filling dependence of the total chemical potential. **c.** Correlated doping evolution of spectral features at $K_T$ (top) and $\Gamma$ (middle), alongside the inverse compressibility from ref.[19] (bottom). The top and middle panels are zoom-ins of data from panels **a** and **b** near charge neutrality. The flat band at $K_T$ (associated with *f*-electrons) shifts downward with increasing $\nu$, reaching the Fermi level near $\nu \approx 0.6$, where it begins to fill (black dashed line). Simultaneously, the Dirac point of the dispersive band at $\Gamma$ (associated with *c*-electrons) also shifts downward and then reverses, moving back up as the f-band becomes populated. This reversal is nicely correlated with the total inverse compressibility becoming negative, signaling the reshuffling of charge from c- to f-electrons. **d.** Schematic depiction of the energy levels of *c*- and *f*-electrons as a function of filling, ignoring hybridization for simplicity. At $\nu = 0$, the *f*-electrons exhibit upper and lower Hubbard bands, split by the on-site Coulomb energy, $U$, while the *c*-electrons show a Dirac-like dispersion. At $0 < \nu \lesssim 0.6$, the c-electrons are gradually filled, and the f-electrons bands are shift down toward the Fermi energy. For $0.6 \lesssim \nu < 1$, the *f*-bands reaches the Fermi energy and get gradually populated, while the c-electron Dirac point moves upward toward the Fermi energy. This is the origin of the observed "Dirac revivals"[19] . At $\nu \approx 1$, as the f-electron band becomes full, a new *f*-electron "Hubbard band" emerges at an energy $U$ above, repeating the cycle.

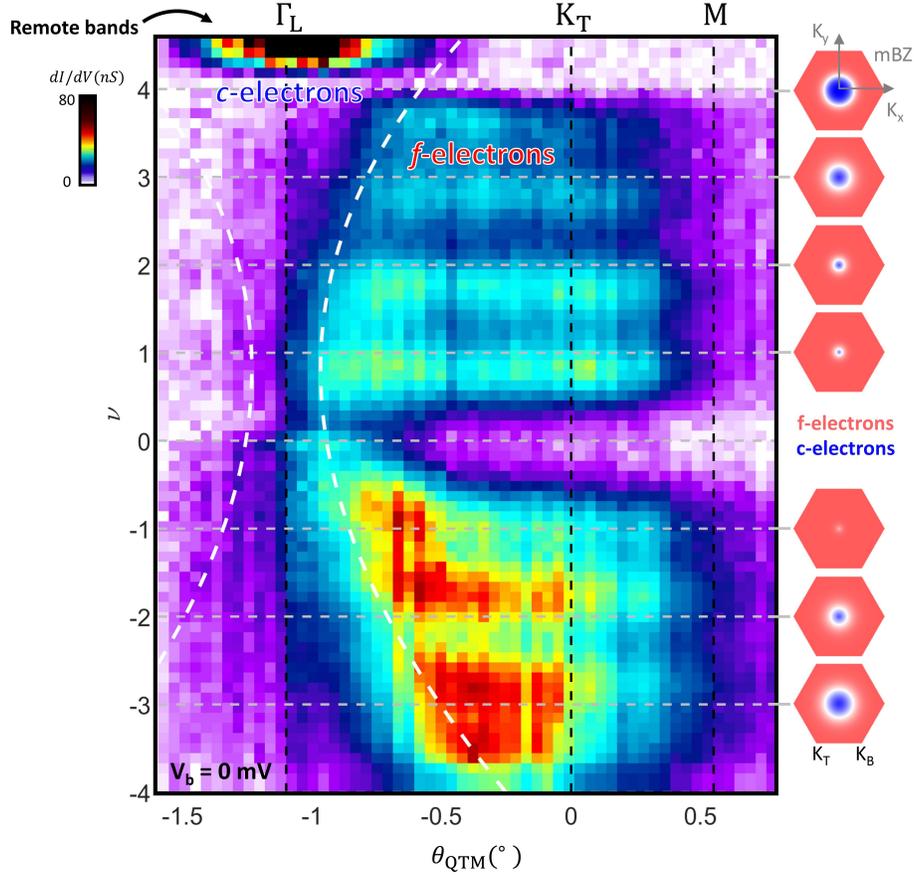

**Figure 5: Doping evolution of the relative fraction of *c*- and *f*- electrons in momentum space.** $dI/dV$ vs. $\theta_{QTM}$ and $\nu$ measured at $V_b = 0mV$. This signal is proportional to the spectral weight at the Fermi energy and reveals where electronic states reside in momentum space as a function of doping. The momenta $\Gamma$, $K_T$, M are labeled on the top x-axis. Integer fillings are indicated by horizontal gray lines. Note that we plot the raw $dI/dV$ signal, whose magnitude drops sharply left to the $\Gamma$ point due to tunneling matrix elements. At charge neutrality ($\nu = 0$), the spectral weight is concentrated near $\Gamma$, and is absent elsewhere. With doping, flat features emerge across a wide momentum range, reflecting the spectral weight of the flat f-electrons bands appearing at the Fermi level. This weight oscillates with a periodicity $\Delta\nu \approx 1$, consistent with the cascades of these bands. As the flat-band features approach the $\Gamma$ point their spectral weight is diminished and transfers to the dispersive *c*-electrons lying below the Fermi energy. The white dashed line traces the momentum boundary at which the $dI/dV$ intensity falls to 80% of its value at a given doping (Methods M8), roughly delineating the boundary between *c*- and *f*- electrons dominated regions. This boundary is mirrored about the $\Gamma$ point, so together the two dashed lines mark the region in momentum space occupied by *c* electrons. The right insets are schematic representation of the *c*- and *f*- electron momentum space regions in the mini-BZ as derived from the measurement in the main panel. The colormap (red to blue) denotes the relative f- and c-electron character. As doping increases (for both electrons and holes), the c-electron momentum-space fraction grows due to the buildup of a Hartree potential that is less repulsive for c-electrons.

**Acknowledgements:** We thank Eva Andrei, Leni Bascones, Andrei Bernevig, Leonid Glazman, Eslam Khalf, Paco Guinea, Pablo Jarillo-Herrero, Maria Jose Calderon, Patrick Ledwith, Allan MacDonald, Felix von-Oppen, Yuval Oreg, Ady Stern, Roser Valenti, Ashvin Vishwanath, Yuval Waschitz and Ali Yazdani for fruitful discussions. Work was supported by the Leona M. and Harry B. Helmsley Charitable Trust grant, the Rosa and Emilio Segre Research Award, the ERC-Adg grant (QTM, no. 101097125), the DFG funded project Number 277101999 - CRC 183 (C02), SNF Sinergia project Nr. CRSII_222792 / 1, BSF grant (2020260) and ISF Quantum grant (1621/24). A.I was supported by the Azrieli Fellow Program. E.B was supported by the European Research Council (ERC) under grant HQMAT (Grant Agreement No. 817799) and CRC 183 of the Deutsche Forschungsgemeinschaft (Project C02).

**Author Contributions:** J.X., A.I., J.B and S.I. designed the experiment., J.X., A.I., and J.B built the cryogenic QTM microscope, fabricated the devices with N.G and Y.Z., and performed the experiments. J.X., A.I., J.B., E.B. and S.I. analyzed the data. K.W. and T.T. supplied the hBN crystals. J.X, A.I., J.B. and S.I. wrote the manuscript with input from other authors.

**Data availability:** The data shown in this paper is provided with this paper. Additional data that supports the plots and other analysis in this work are available from the corresponding author upon request.

**Competing interests:** The authors declare no competing interests.

**Correspondence and requests for materials** should be addressed to shahal.ilani@weizmann.ac.il